\documentclass[preprint,showpacs,showkeys,preprintnumbers,amsmath,amssymb]{revtex4}

\usepackage{graphicx}
\usepackage{dcolumn}
\usepackage{natbib}
\usepackage{color,calc}
\usepackage{bm}

\newcommand{\ybco}{YBa$_2$Cu$_3$O$_{7-\delta}$}

\newcommand{\lsco}{La$_{2-x}$Sr$_x$CuO$_{4}$}
 
\begin{document}

\title{Low-$q$ dispersion of transverse acoustic phonons in \ybco~}
\author{Andrea Gauzzi$^1$}
\email{gauzzi@imem.cnr.it}
\author{Matteo d'Astuto$^2$}
\author{Francesca Licci$^1$}
\author{Alexei Bossak$^3$}
\author{Michael Krisch$^3$}
\affiliation{$^1$IMEM-CNR, Area delle Scienze, 43100 Parma, Italy}
\affiliation{$^2$Physique des Milieux Condens\'es, CNRS-UMR 7602, 
Universit\'e Pierre et Marie Curie, B77\\4, place Jussieu, 
75252 Paris CEDEX 05, France} 
\affiliation{$^3$European Synchrotron Radiation Facility, BP 220, 
38043 Grenoble, France}

\date{\today} 

\begin{abstract}
High-resolution ($\Delta E$=1.3 meV) inelastic X-ray scattering (IXS) in optimally doped untwinned \ybco~ single crystals show that the dispersion of transverse acoustic phonons in the $a$-direction remains linear with no softening due to bending modes in the long wavelength limit, down to $q$=0.02. The sound velocity is found to be 2700 $\pm$ 70 m s$^{-1}$, in excellent agreement with ultrasound measurements, and unchanged upon cooling from room temperature to 98 K. These results rule out a strong coupling between bending dynamics of the CuO$_2$ planes and charge carriers in cuprates.
\end{abstract}

\pacs{74.72.Bk, 74.25.Kc, 78.70.Ck}

\keywords{Y-based cuprates, phonons, X-ray scattering.}

\maketitle

Recent experimental studies renewed the interest in the possible role of lattice dynamics in the superconducting state of cuprates. In particular,  Inelastic Neutron \cite{mcq,pin3} and X-ray \cite{das} Scattering (INS, IXS) studies support early evidence \cite{pyk} of ubiquitous phonon instabilities in the superconducting CuO$_2$ planes at commensurate in-plane wavevectors \textbf{q}. Further progress in the field requires to establish whether these instabilities affect, or merely coexist with, the superconducting state. On one hand, it has been established that lattice instabilities are common in perovskite-like compounds, independent of superconductivity \cite{zho}. A coexistence between these instabilities and superconductivity is supported by a recent study on short-range oxygen orderings in \ybco~ \cite{str}. The authors found that incipient lattice modulations are independent of electronic doping and, hence, of the superconducting critical temperature, $T_c$. In \lsco~, INS data support a picture of incipient stripe order phase with the same periodicity in superconducting and non-superconducting samples \cite{pin2}. On the other hand, previous theoretical studies indicate that finite $q$ anomalies, a signature of incipient lattice modulations, may compete \cite{bec} with superconductivity but also enhance \cite{all} the electron-phonon mediated pairing, provided the modulations are dynamic and the metallic state preserved.

In order to clarify this point, in this work we consider a new type of lattice instabilities corresponding to bending modes, with $\omega, q \approx 0$. As we shall discuss below, these modes are potentially interesting for enhancing $T_c$, whenever a sizeable softening of these modes is concomitant to a strong coupling with the charge carriers. In cuprates, as well as in most systems, the $q \approx 0$ region of the phonon dispersions was extensively investigated for the optical modes, but remains unexplored for the acoustic ones \cite{pin1}. To our knowledge, this incompleteness of available data is due to the experimental difficulty of combining a high resolution in energy, $E=\hbar \omega$, with a high resolution in momentum transfer, $q$. Thus, an adequate experimental set up requires simultaneously a highly monochromatic and collimated neutron or X-ray beam, an energy analyzer with high energy and momentum resolution and high-quality single crystals with a low degree of mosaicity. To explore the above region of phonon dispersion for acoustic modes would be important to gain a full insight into the complex dynamics of cuprates, thus enabling more stringent tests of competing theories of high-$T_c$ superconductivity. 

In this context, the in-plane transverse acoustic (TA) mode is particularly interesting. For this mode, a bending dynamics with a softening at sufficiently low $q \lesssim 0.1 $ in reduced lattice units (r.l.u.) is expected in layered compounds and would be favored by the covalent in-plane Cu-O bonds of cuprates. According to the standard theory of elasticity in continuous media \cite{lan1}, assuming isotropic layers, the expected dispersion relation is as follows:
\begin{equation}
\omega^2=U_\parallel^2q_\parallel^2+U_\perp^2q_\perp^2+\gamma q_\parallel^4
\label{eq:bending}
\end{equation}
where $\omega$ is the wave frequency, $\gamma$ is the bending stiffness constant of the layer, the suffixes $\parallel$ and $\perp$ indicate the in- and out of-plane quantities, $U=c_{44}/\varrho$ would be the square of the velocity of sound in the absence of the last term, $c_{44}$ is the layer shear elastic constant and $\varrho$ is the volume density. A bending-induced softening of in-plane TA modes in agreement with the above expression was experimentally observed in graphite \cite{nic}, graphite intercalated compounds \cite{zab}, and layered transition metal dichalcogenides like 2H-MoS$_2$ \cite{wak}. Such peculiar lattice dynamics greatly enhances the electron-phonon interaction without invoking any non-linear phenomena. Taking into account Eq. (1), in first approximation, the vertex part of the interaction takes the following form \cite{lan2}:
\begin{equation}
\Gamma(q,\omega \approx 0) \approx \frac{w^2}{\varrho U^2}\frac{1}{1+\frac{\gamma}{U^2}q^2+i\epsilon}
\label{eq:vertex}
\end{equation}
where $w$ expresses the strength of the electron-phonon coupling, $\epsilon$ is an arbitrarily small quantity and the other symbols are as above. From Eq. (\ref{eq:vertex}), it is apparent that the vertex tends to diverge at small $q$ values for large $\gamma/U^2$ ratios, \textit{i.e.} if the mode energy tends to zero, as experimentally found in the aforementioned layered compounds where the bending (or shear) dynamics dominates. This situation may be relevant to the superconducting state of cuprates, since early experimental studies put into evidence a strong coupling between the superconducting order parameter and shear lattice distortions in \lsco~ and \ybco~ \cite{bha}. To verify this possibility, in the present work we searched for a possible softening of the in-plane TA mode in the hitherto unexplored low-$q$ region. Using IXS with a very high resolution in energy, $\Delta E$=1.3 meV, we measured the dispersion relation from $q$=0.3 down to $q$=0.02 in \ybco~ single crystals. In the $q \leq 0.1$ region, the data show no deviation from linearity either at room temperature or at 98 K, the latter temperature being well inside the region of superconducting fluctuations. These results rule out a strong coupling between bending dynamics of the CuO$_2$ layers and high-$T_c$ superconductivity in cuprates, and furthermore demonstrate the feasibility of accurate measurements of very low-$q$ modes by means of IXS.

The experiment was carried out on the undulator ID28 beamline at the ESRF. We employed beam spot sizes as small as $\approx 0.2 \times$0.1 mm$^2$ and two different Si (11 11 11) and (13 13 13) monochromator Bragg reflections \cite{mono}. Using the ID28 6m-long arm spectrometer, these reflections provided an overall energy resolution of 1.7 and 1.3 meV, respectively \cite{ana,id28}. We investigated the in-plane TA dispersion along $\Delta$ ($a$ or [$\xi$,0,0] direction) with out-of-plane ([0,0,$\zeta$]) polarization. One can verify that the $q$ resolution for this phonon, $\Delta Q_a$, is limited by two independent factors: 1) the mosaic spread of the crystal, $\Delta \vartheta_c$, convoluted with the divergence, $\Delta K_i$, of the incident beam, the convolution being measured experimentally by the rocking curve width, $\Delta \omega$; 2) the transverse component, $\Delta2\theta_t$, of the width, $\Delta2\theta$, of the scattering solid angle. In a typical INS experiment, $\Delta Q$ is usually dominated by the former factor. In the present IXS experiment, one can significantly enhance $\Delta Q_a$ thanks to the very small beam divergence (120 $\mu$rad horizontal $\times$ 60 $\mu$rad vertical) characteristic of the synchrotron undulator sources and optimized beamline optics, provided a high quality crystal with a sufficiently low degree of mosaicity is used. This is the case of the present experiment. We measured two \ybco~ single crystals grown by the flux method, as described elsewhere \cite{lic}. The crystals are small platelets of about $\sim$ 1 $\times$ 1 $\times$ 0.1 mm$^3$ size selected from two batches prepared under nominally the same conditions. Oxygen post-annealing at 500 $^\circ$C for two weeks enabled to achieve a full-oxidation ($\delta \approx 0.1$). The degree of mosaicity was very low in both crystals, as indicated by the small full widths at half maximum of the (006) rocking curves, $\Delta\omega$=0.009-0.018$^\circ$ for both samples, that are close to the horizontal beam divergence, 0.007$^\circ$. This gives a mosaic spread of 0.005-0.016$^\circ$. Using the Si (13 13 13) reflection, with $\lambda_i$=0.4824 \AA, it turns out that the contribution of $\Delta\omega$ to $\Delta Q$ is in the 0.0003-0.0006 r.l.u. range in the (0,0,6) Brillouin zone. We used two configurations of horizontal slits with horizontal openings of 0.190 and 0.095$^\circ$, the latter for the measurements at $q$ points near the zone center. To minimize the spurious contribution from the (0,1,0) direction near the zone center, the opening of the vertical slits of the analyzer was reduced to decrease the sagittal scattering angle from 0.57 to 0.19$^\circ$. A straightforward calculation shows that, under the above experimental conditions, $\Delta Q$ along the longitudinal, $[\xi,0,0]$, and transverse, $[0,0,\zeta]$, directions in the (0,0,6) Brillouin zone is in the 0.001-0.002 and 0.04-0.08 r.l.u. range respectively. Hence, in the present experiment, $\Delta Q$ is dominated by the divergence of the scattered beam, \textit{i.e.} by the analyzer slit opening. In the case of the Si (11 11 11) reflection, with $\lambda_i$ = 0.5701 \AA, the estimated $\Delta Q$ values are increased by an amount comparable with the 16\% increase of $\lambda_i$ of this reflection with respect to the (13 13 13) one.
 \begin{figure}
\includegraphics[scale=0.4]{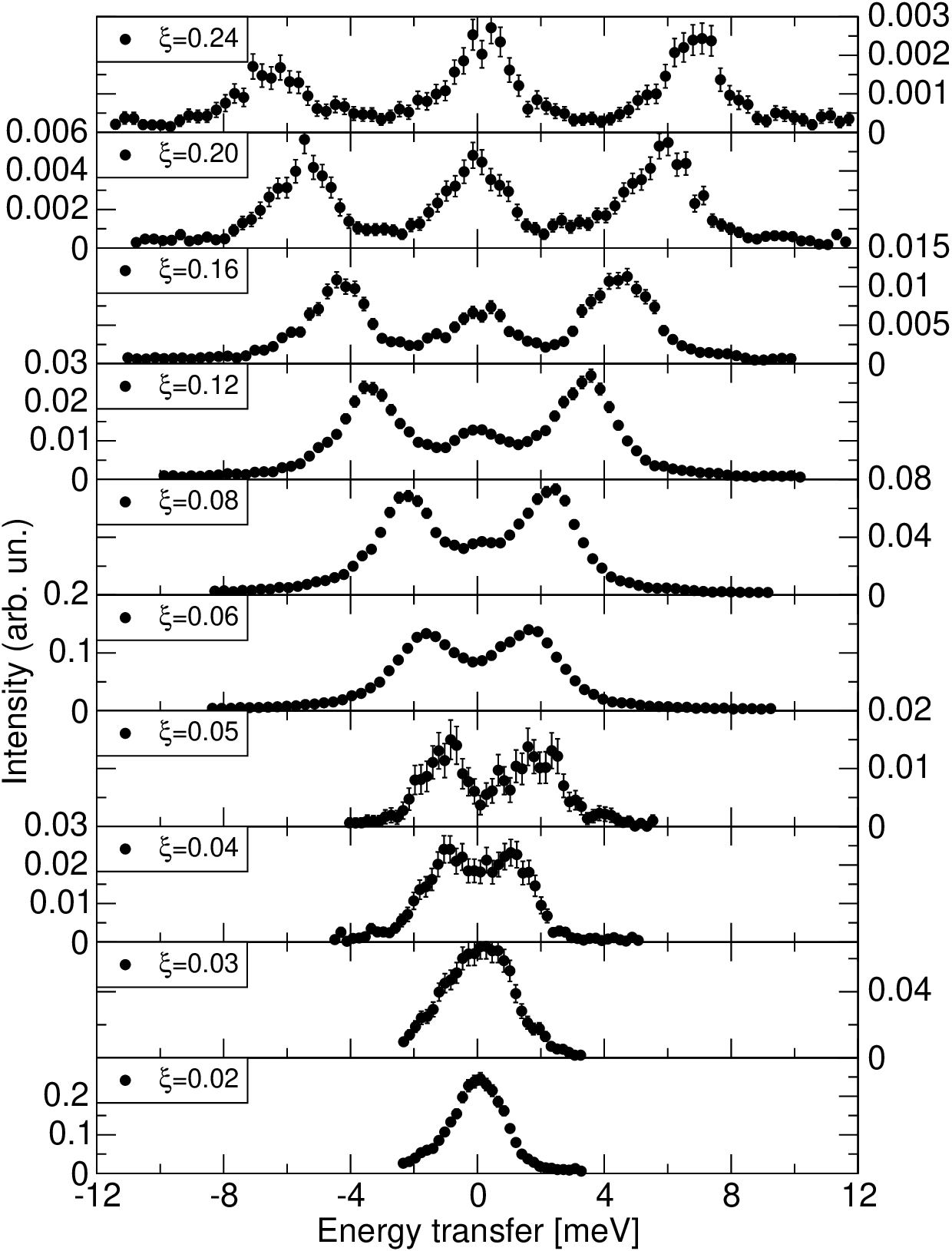}
\caption{\label{scans} Energy scans near (0,0,6) for various values of the in-plane momentum transfer $\xi$ along the $a$-axis (indicated in the legend) for one of the two \ybco~ single crystals. These data were obtained using the Si (13 13 13) reflection, which provides an overall energy resolution of 1.3 meV.}
\end{figure}

\begin{figure}
\includegraphics[scale=0.4]{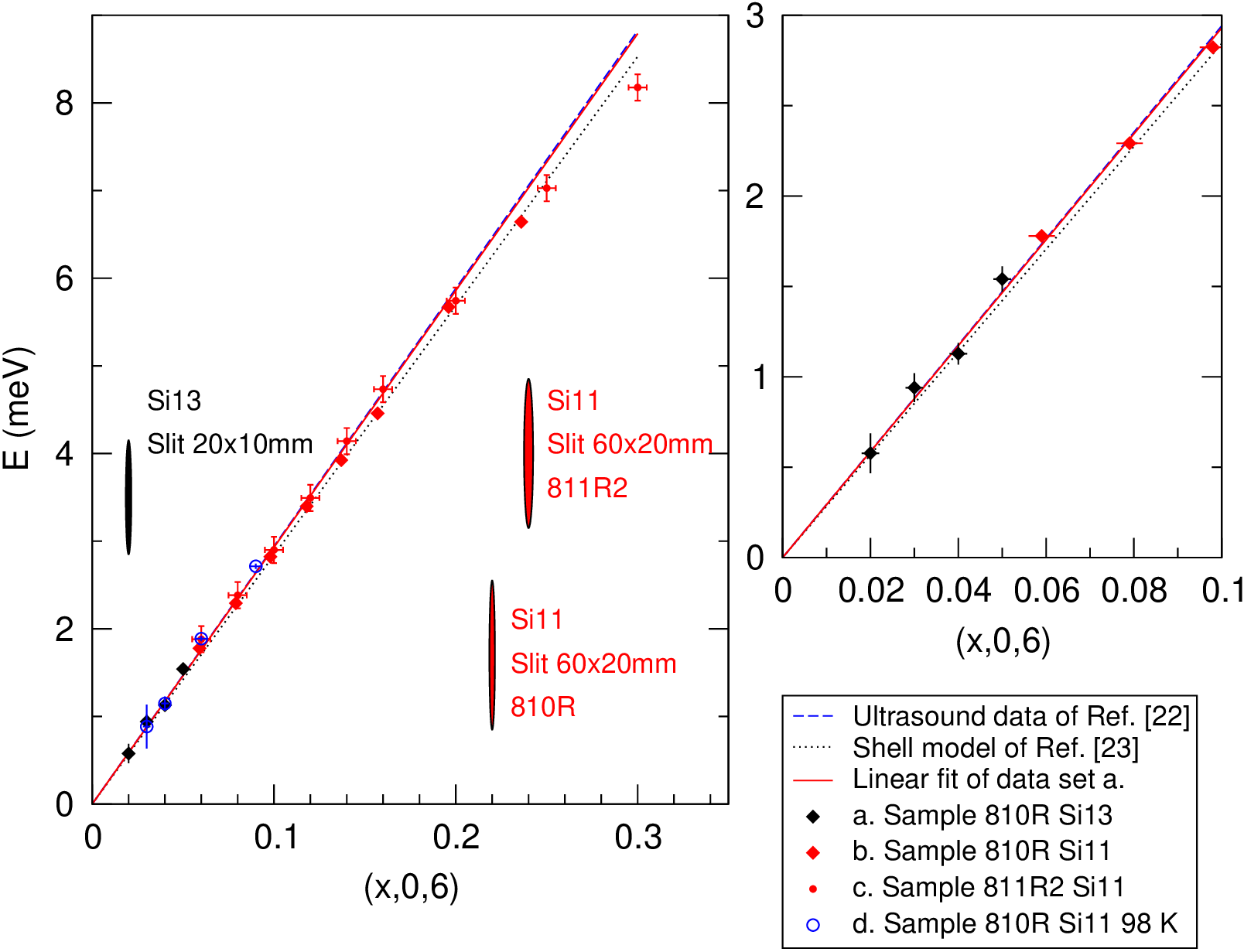}
\caption{\label{disp} Dispersion relation of the in-plane TA phonon at room temperature and 98 K for the two \ybco~ crystals labelled as 810R and 811R. The right panel shows in detail the very low $q$ data for the former. Vertical lines indicate the energy and momentum resolutions that depend upon the Si (11 11 11) or (13 13 13) reflection used, slit openings and crystal mosaicity (see text). }
\end{figure}

The phonon dispersion was measured in a fixed-$Q$ configuration at various ($\xi$,0,6) points of the Brillouin zone in the vicinity of the (006) Bragg reflection. The latter reflection is the most intense one of the (00$l$) type in fully oxygenated \ybco~, thus providing the largest photon-phonon scattering cross-section for low-$q$ acoustic phonons. The same runs of measurements were carried out on the two crystals. In Fig. \ref{scans}, a series of room temperature energy scans at various values of phonon momentum, $\xi$, is reported. For all $\xi$ values, one clearly distinguishes the triple peak structure consisting of the central peak, that corresponds to diffuse elastic scattering, and of the two lateral phonon, Stokes and anti-Stokes, peaks. The Si (11 11 11) data enabled to resolve the phonon peaks down to $\xi =0.04$. At lower $\xi$ values, these peaks appeared as too small shoulders of the central peak and could not be resolved. To resolve them, it was necessary to use the (13 13 13) reflection and closer analyzer slits, as described above. The relatively low intensity of the central peak as compared to the phonon peaks indicates that the amount of disorder, responsible for the diffuse scattering, is limited in the crystal.
\begin{figure}
\includegraphics[scale=0.4]{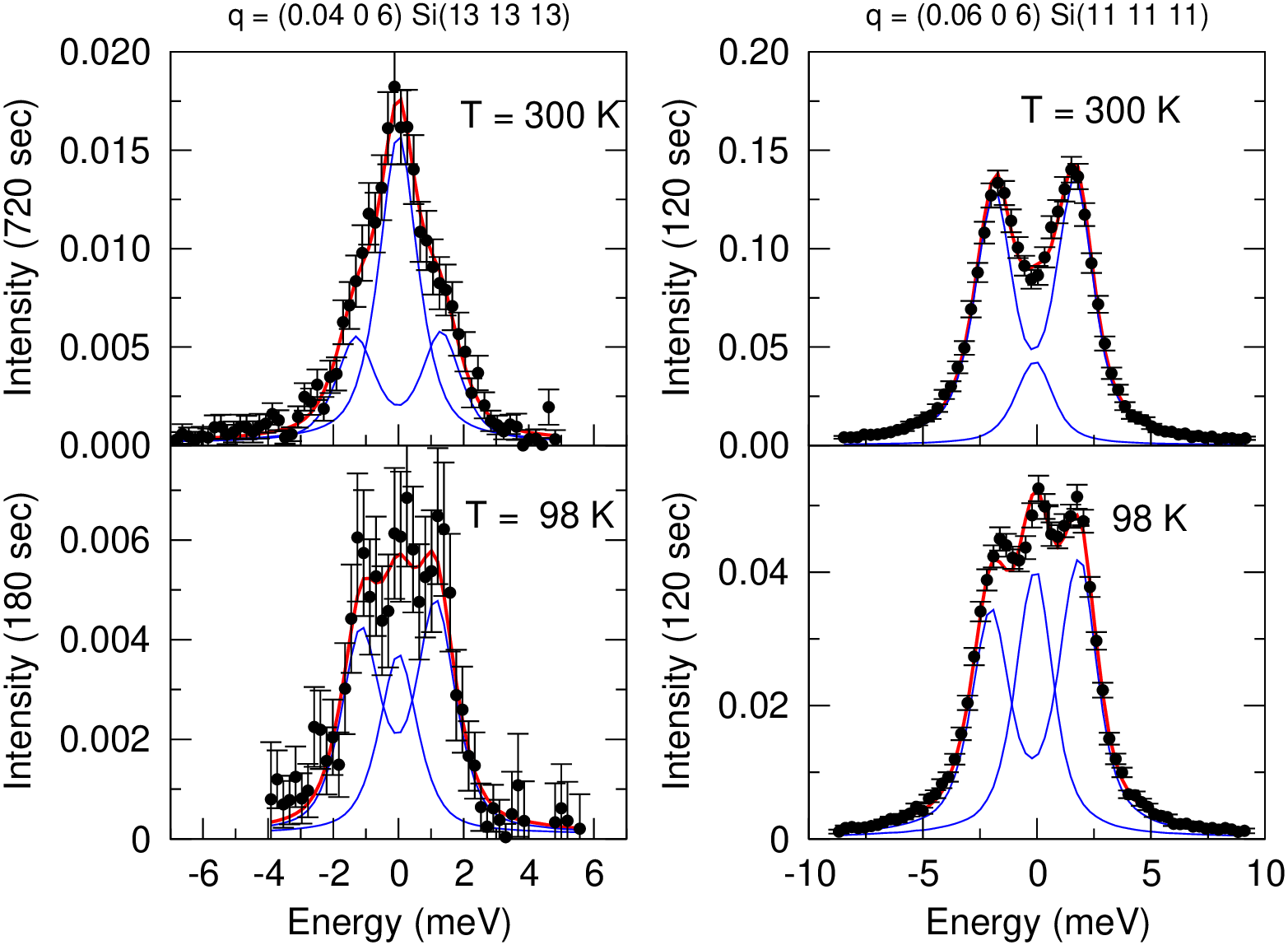}
\caption{\label{scantdep} Comparison between the TA phonon peaks at 300 and 98 K measured using the Si (11 11 11) and (13 13 13) reflections. Solid lines are a fit of the data as described in the text.}
\end{figure}
For every energy scan, the above triple peak structure was fitted using the standard constraints regarding the relative position and intensity of the two lateral Stokes and anti-Stokes peaks set by the Bose-Einstein statistics. The phonons peaks and the diffuse elastic scattering peak were modelled by harmonic oscillators and a delta function respectively. The sum of these contributions was convoluted with the instrumental resolution function as described in Refs. \onlinecite{das,hgixs}. Four examples of fitted energy scans are shown in Fig. \ref{scantdep}. The agreement within the error bars between the experimental data and the model is evident. 
The phonon energy values extracted from all scans yield in a straightforward manner the dispersion reported in Fig. \ref{disp}, where the results obtained at room temperature and 98 K are summarized. One notes that the data of the two samples are reproducible within the experimental uncertainty. The accurate IXS measurements enable us to firmly conclude that the dispersion is linear down to $q$=0.02. A linear regression in the $q\leq$ 0.1 range of the data of sample 810R, which gave the most accurate results thanks to the lower mosaicity, yields a transverse sound velocity along the $a$ direction of 2700 $\pm$ 70 m s$^{-1}$. This value is in excellent agreement with the value of 2710 m s$^{-1}$ derived from a previous ultrasound measurement of $c_{55}$ \cite{ult}, the relevant elastic constant for orthorhombic \ybco~. Good agreement is also found with the value of 2620 m s$^{-1}$ obtained from a shell model calculation of $c_{55}$ \cite{cha}. The discrepancy in the latter case is compatible with the approximations of the model. We finally fitted the same data set using a sine function to take into account the levelling off of the mode frequency at the zone boundary. This fit yields a slightly higher velocity, 2810 m s$^{-1}$, which suggests that the linear fit may underestimate the slope of the dispersion curve even at $q\leq$ 0.1. In any case, it is apparent from Fig. \ref{disp} that the present measurements agree with the ultrasound ones within the error bars. 

Neither a hardening nor a softening of the mode is found upon cooling from room temperature down to 98 K, as it is noted from Fig. \ref{disp} and, in more detail, Fig. \ref{scantdep}. It is recalled that the latter temperature is well inside the region of thermodynamic fluctuations of the superconducting order parameter in \ybco~. It follows that no change of the lattice dynamics regarding the in-plane TA mode occurs in presence of the superconducting instability. By assuming the relevance of the electron-phonon interaction to the pairing mechanism in cuprates, one concludes that this mode would not be strongly involved in the pairing process. At best, according to a well established result of the BCS theory \cite{axe}, a renormalization of the phonon frequency and lifetime is expected at lower temperatures as a general consequence - rather than a cause - of the superconducting transition that affects all phonon modes.

In conclusion, using IXS with very high resolution in energy and momentum, we performed the first measurement of in-plane TA phonon dispersion in the very low $q$-limit on cuprates. Stringent conditions regarding the monochromaticity and the collimation of the beam combined with the low degree of mosaicity of the samples enabled us to measure reproducibly the dispersion along the $a^*$-axis in two untwinned \ybco~ single crystals. Upon cooling from room temperature to 98 K, well inside the region of superconducting fluctuations, no anomaly was found down to $q= 0.02$, the dispersion remains linear within the experimental uncertainty and the corresponding sound velocity is unchanged. Its value is in agreement with previously reported ultrasound measurements and shell model calculations. The absence of any anomalies or temperature-induced mode hardening or softening in the region of superconducting fluctuations indicates that, in cuprates, the bending dynamics and the coupling between in-plane TA acoustic modes and the electronic degrees of freedom are negligible. This result should set stringent conditions for any phonon-based model of high-$T_c$ superconductivity in these materials. In addition, our experiment shows the suitability of IXS to accurately measure low-energy excitations in the $q$=0 limit, provided single crystal with a low degree of mosaicity are used. 

This work has been supported by the European Synchrotron Radiation Facility (Experiment HS-2445). The authors acknowledge the technical assistance of T. Besagni and P. Ferro.


\begin {thebibliography}{100}

\bibitem{mcq} R. J. McQueeney, Y. Petrov, T. Egami, M. Yethiraj, G. Shirane, and Y. Endoh, Phys. Rev. Lett. \textbf{82}, 628 (1999).
\bibitem{pin3} L. Pintschovius, D. Reznik, and W. Reichardt, Y. Endoh, H. Hiraka, J. M. Tranquada, H. Uchiyama, T. Masui, and S. Tajima, Phys. Rev. B {\bf 69}, 214506 (2004).
\bibitem{das} M. d'Astuto, P. K. Mang, P. Giura, A. Shukla, P. Ghigna, A. Mirone, M. Braden, M. Greven, M. Krisch, and F. Sette, Phys. Rev. Lett. \textbf{88}, 167002 (2002).
\bibitem{pyk} N. Pyka, W. Reichardt, L. Pintschovius, G. Engel, J. Rossat-Mignod, and J.Y. Henry, Phys. Rev. Lett. {\bf 70}, 1457 (1993).
\bibitem{zho} W. Zhong and D. Vanderbilt, Phys. Rev. Lett. {\bf 74}, 2587 (1995).
\bibitem{str} J. Strempfer, I. Zegkinoglou, U. R\"{u}tt, M. v. Zimmermann, C. Bernhard, C. T. Lin, Th. Wolf, and B. Keimer, Phys. Rev. Lett. {\bf 93}, 157007 (2004).
\bibitem{pin2} L. Pintschovius and M. Braden, Phys. Rev. B {\bf 60}, R15039 (1999).
\bibitem{bec} F. Becca, M. Tarquini, M. Grilli, and C. Di Castro, Phys. Rev. B {\bf 54}, 12443 (1996). 
\bibitem{all} P. B. Allen in \textit{Dynamical Properties of Solids} vol. 3, Eds. G.K. Horton and A.A. Maradudin (North-Holland, Amsterdam, 1980) and references therein.
\bibitem{pin1} See, for example, L. Pintschovius and W. Reichardt in \textit{Physical Properties of High Temperature Superconductors}, ed. D. Ginsberg (World Scientific, Singapore, 1995) and references therein.
\bibitem{lan1} L. D. Landau and E. M. Lifshitz, in \textit{Theory of Elasticity} 3$^{rd}$ ed. (Pergamon Press, Oxford, 1986).
\bibitem{nic} R. Nicklow, N. Wakabayashi, and H. G. Smith, Phys. Rev. B {\bf 5}, 4951 (1972).  \bibitem{zab} H. Zabel, J. Phys.: Condens. Matter {\bf 13}, 7679 (2001).
\bibitem{wak} N. Wakabayashi, and R. Nicklow in \textit{Electrons and Phonons in Layered Crystal Structures} Eds. T. J. Wieting and M. Schl\"{u}ter (Reidel, Dordrecht, 1979).

\bibitem{lan2} L. D. Landau and E. M. Lifshitz, in \textit{Statistical Physics, Part 2} (Pergamon Press, Oxford, 1981).
\bibitem{bha} S. Bhattacharya, M. J. Higgins, D. C. Johnston, A. J. Jacobson, J. P. Stokes, D. P. Goshorn, and J. T. Lewandowski, Phys. Rev. Lett. {\bf 60}, 1181 (1988).
\bibitem{mono} R. Verbeni, F. Sette, M. H. Krisch, U. Bergmann, B. Gorges, C. Halcoussis, K. Martel, C. Masciovecchio, J.F. Ribois, G. Ruocco, H. Sinn, J. Synchrotron Radiation {\bf 3}, 62 (1996).
\bibitem{ana} C. Masciovecchio, U. Bergmann, M. Krisch, G. Ruocco, F. Sette, R. Verbeni, Nucl. Instr. and Meth. B {\bf 111}, 181 (1996); C. Masciovecchio, U. Bergmann, M. Krisch, G. Ruocco, F. Sette, R. Verbeni, Nucl. Instr. and Meth. B {\bf 117}, 339 (1996).
\bibitem{id28} M. Krisch, R. A. Brand, M. Chernikov and H. R. Ott, \textit{Phys. Rev. B} \textbf{65}, 134201 (2002).
\bibitem{lic} F. Licci, C. Frigeri, and H. J. Scheel, J. of Crystal Growth {\bf 112}, 606 (1991).
\bibitem{hgixs} M. d'Astuto, A. Mirone, P. Giura, D. Colson, A. Forget, and M. Krisch, J. Phys.: Condens. Matter \textbf{15}, 8827 (2003).
\bibitem{ult} M. Lei, J. L. Sarrao, W. M. Visscher, T. M. Bell, J. D. Thompson, A. Migliori, V. W. Welp, B. W. Veal, Phys. Rev. B \textbf{47}, 6154 (1993).
\bibitem{cha} S. L. Chaplot, W. Reichardt, L. Pintschovius, N. Pyka, Phys. Rev. B \textbf{52}, 7230 (1995).
\bibitem{axe} J. D. Axe and G. Shirane, Phys. Rev. Lett. \textbf{30}, 214 (1973); S. M. Shaprio, G. Shirane, and J. D. Axe Phys. Rev. B \textbf{12}, 4899 (1973).
\end {thebibliography}
\end{document}